\documentclass[aps,prl,twocolumn,superscriptaddress,showpacs]{revtex4}
\usepackage{epsfig}
\usepackage{amsmath}
\usepackage{times}

\begin{document}

[Physical Review E, to appear]

\title{Signatures of small-world and scale-free properties in large computer programs}

\author{Alessandro P. S. de Moura}
\affiliation{Instituto de F\'{\i}sica, Universidade de S\~{a}o Paulo,
Caixa Postal 66318, 05315-970 S\~{a}o Paulo, Brazil}

\author{Ying-Cheng Lai}
\affiliation{Department of Mathematics and Statistics,
Arizona State University, Tempe, Arizona 85287}
\affiliation{
Department of Electrical Engineering, Department of Physics, 
Arizona State University, Tempe, Arizona 85287}

\author{Adilson E. Motter}
\affiliation{Department of Mathematics and Statistics,
Arizona State University, Tempe, Arizona 85287}

\date{\today}
 
\begin{abstract}
A large computer program is typically divided into many hundreds or
even thousands of smaller units, whose logical connections define a
network in a natural way. This network reflects the internal structure
of the program, and defines the ``information flow'' within the
program.  We show that, (1) due to its growth in time this network
displays a {\it scale-free} feature in that the probability of the
number of links at a node obeys a power-law distribution, and (2) as a
result of performance optimization of the program the network has a
{\it small-world} structure.  We believe that these features are
generic for large computer programs.  Our work extends the previous
studies on growing networks, which have mostly been for physical
networks, to the domain of computer software.
\end{abstract}
\maketitle  

Large computer programs nowadays are becoming increasingly more complex. 
Such a program can easily contain hundreds of thousands or even millions
of lines of code. In order to make the programs manageable, the code
is split into many small files that are linked together in a coherent
but quite sophisticated fashion. A large computer program can thus
be regarded as a complex network. But what are the characteristics
of such a network?

Some basic features about large computer programs are the following.
First, they are {\it dynamic} in that they continue to evolve in time.
For instance, the beginning versions of a program may be relatively
simple and small in size. As time goes, application demand increases,
resulting in continuous expansion of the program in many
aspects. Thus, the underlying networks may be regarded as {\it growing
networks}. Second, there exists a number of ``key'' components of the
program which are linked to many other components (such as
subroutines). As new components developed for new applications are
added to the program, they are more likely to be linked to the key
components of the program. That is, the network develops according to
the rule of {\it preferential attachment}.  As argued by Barab\'{a}si
{\it et al.} in their seminal work \cite{BA:1999,BAJ:1999}, growth with
preferential attachment is one possible dynamical mechanism
responsible for the network to exhibit the scale-free characteristic,
i.e., a power-law scaling for the probability distribution of the
number of links at a node.

For a dynamically growing network, however,
at a given time, one can also view it as ``static'' and ask for
the topology of the connections between the nodes. Most networks occurring
in nature are large, as they usually contain a huge number of nodes, but they are
sparse in the sense that the average number of links per node is typically much
less than the total number of nodes. Sparse networks
can be characterized as {\it regular}, {\it random}, and {\it small world}. Most regular networks
possess the property that if two nodes are connected to a common third node, then there
is a high probability that the two nodes are connected between themselves. That is,
the network has a high degree of {\it clustering}. However, in general it takes
many steps to move between two arbitrary nodes in the network, i.e., the {\it shortest
possible path} to go from one node to another can be long (in a statistical sense).
A high degree of clustering and a large value for the average shortest path
are thus the two defining properties of most locally connected regular networks. At the opposite end
are random networks \cite{Bollobas:book}: 
due to the sparsity and random connections, such  networks have
extremely low degree of clustering and small average shortest path. Regular
and random networks had been the main focus of research on
network structure and dynamics. It was
pointed out in Ref. \cite{WS:1998} that there exists a physically realizable
range of network topology for which the degree of clustering can be almost as
high as that of a regular network, but the average shortest path can be almost as small
as that of a random network. These are small-world networks. Structurally,
a small-world network differs from a regular one in that there exist a
few random links between distant nodes in the former. Watts and Strogatz argued
that the small-world configuration is expected to be found commonly in large,
sparse networks of the real world. Indeed, examples of small-world networks identified
so far occur in almost every branch of science, which include nervous system
\cite{WS:1998,SHBNYK:2000}, epidemiological invasions \cite{Keeling:1999,LM:2001},
business management \cite{Kogut:2000}, electrical power grid \cite{watts:book},
Internet and World Wide Web \cite{AJB:1999,AJB:2000,BAJ:2000,Barabasi:2001},
social networks \cite{SP:2000,KW:2001},
metabolism \cite{FW:2000},
scientific-collaboration network \cite{Newman:2001a,Newman:2001b}, Ising model
in physics \cite{Gitterman:2000},
religion and economic growth network \cite{BD:2001}, polymer networks \cite{GB:2001},
gene network \cite{GBBK:2002}, and linguistics \cite{MMLP:2002}. 

In this paper, we investigate the network properties of large computer programs
and present results for four widely used computer programs, whose codes are
publically available and can be downloaded from the Internet. They
are (1) the Linux kernel, the core program of the Linux
operating system; (2) ``Mozilla'', the open source version of the
web-browser Netscape; (3) ``XFree86'', the Unix X-Window graphics
package; and (4) ``Gimp'', an image manipulation program for Unix. 
We study the structure of these programs and develop a natural 
way to construct the networks underlying these programs. We 
provide strong evidence that the networks are scale-free and small worlds. While both the
scale-free and small-world features have been demonstrated in many {\it physical}
(or ``hardware'' type of) networks such as the Internet, the World Wide Web,
and actor collaboration networks \cite{BA:1999,WS:1998,Strogatz:2001,AB:review,N:review},
our work demonstrates, for the first time, that these features also govern the network 
dynamics and topology in the {\it software} domain of computer science.

The programming language of choice for encoding large complex programs
is C (and its offspring C++). 
In order to make a program manageable, the code is split into
many small files. These files are of two kinds: {\em source} files and
{\em header} files. The source files (usually with names terminating
in ``.c'' or ``.cpp'') contain the actual code, whereas the header
files (with termination ``.h'') have definitions of variables,
constants, data structure and other information needed by the source
files. A large program typically consists of thousands of source and
header files. If a source file needs the information contained in a
header file, that file is ``included'' in the source file with an
``\#include'' clause. For example, if the source file ``main.c'' needs
some data structure defined in ``sys.h'', it contains a statement such as
``\#include $<$sys.h$>$'', whereby contents of ``sys.h'' are
made accessible to ``main.c''. 

A network can now be defined from the set of source and header files,
as follows. The nodes of the network are header files, and two nodes
are defined to be connected if the corresponding header files are
both included in the same source file. Connected header files are thus
functionally related (they ``work together'' to help the source file
in which they are both included do its job). 
By using a simple program that automatically scans every source file to
see which header files each one of them includes, we generate the
network corresponding to each of the four large programs aforementioned. We
note that a few header files included in the source files belong to
external libraries, and are not part of the program itself. When
generating the networks, we ignore such files. Also, we only consider
the largest connected component of the network, which includes over
$90\%$ of all nodes in all four cases.

\begin{figure}
\begin{center}
\epsfig{figure=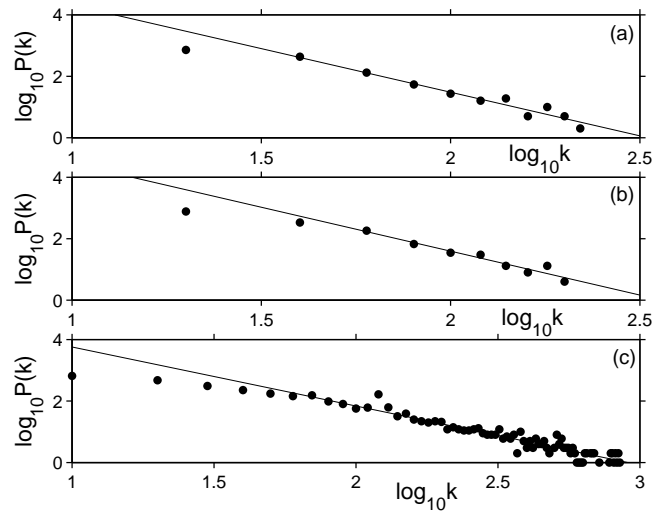,width=\linewidth}
\caption{Algebraic scaling behavior of the (non-normalized) probability  $P(k)$ of the underlying 
networks for widely used computer
programs: (a) the Linux kernel, (b) XFree86, and (c) Mozilla.}
\label{fig:code_scaling}
\end{center}
\end{figure}

We first present results concerning the scale-free feature of the 
computer-code networks. Let $k$ be the variable that measures the number
of links at different nodes in the network. For a network that contains
a large number of nodes, $k$ can be regarded as a random variable. 
Let $P(k)$ be the probability distribution of $k$. A scale-free network
is characterized by the following algebraic scaling behavior in $P(k)$:
\begin{equation} \label{eq:algebraic}
P(k) \sim k^{-\gamma},
\end{equation}
where $\gamma$ is the scaling exponent. As pointed out in Refs.
\cite{BA:1999,BAJ:1999,AB:review}, many real networks, such as the
Internet, the World Wide Web, and the network of movie actors,
appear to be scale-free with the value of the exponent 
ranging from 2 to 3. The theoretical model proposed in Ref. \cite{BAJ:1999}
suggests the following two basic features in the network
dynamics, which determine the algebraic scaling law:
growth and preferential attachment. For growth, one can start
with a small number $m_0$ of vertices and at every time step add a new vertex with 
$m$ edges to the network, where $m \le m_0$. For preferential 
growth, one can choose the probability that a new link is to be added to
the $i$th node to be proportional to the number of links already 
existing in that node. The scaling law (\ref{eq:algebraic}) can be derived from 
these two conditions \cite{BAJ:1999}. 
Fig. \ref{fig:code_scaling} shows the scaling behavior of $P(k)$
for three of the computer programs that we consider here, where (a-c) correspond
to the Linux kernel, XFree86, and Mozilla, respectively. [The total number
of nodes in the network associated with Gimp is too small to allow for
the statistical quantity $P(k)$ to be computed.] For  large $k$, a robust
algebraic scaling behavior is present in all the three cases, where the scaling
exponents are $\gamma_{Linux} \approx 2.8$, $\gamma_{XFree86} \approx 2.9$,
and $\gamma_{Mozilla} \approx 1.9$. These results
suggest that large computer programs can be regarded as scale-free,
growing networks \cite{Exponent}.

We next turn to the small-world feature of the large computer-program networks.
For a given program, once the underlying network is built up, we can
calculate the quantities that 
characterize their statistical properties; these are shown in Table I,
for each program. We see that the average number of links per node
$\mu$ in all networks is much smaller than the total number of nodes
$N$, which means that the networks are {\em sparse}, a necessary
condition for the notion of small-world network to be meaningful. The
quantities of interest to us are the {\em average shortest path} $L$,
which is the average over all pairs of nodes of the number of links in
the shortest path connecting the two nodes; and the {\em clustering}
$C$, which is the probability that two nodes $a$ and $b$ are
connected, given that they are both connected to a common third node $c$. If
$C$ is close to 0, the network is not locally structured; if $C$ is close to 1,
the network is highly clustered.

\begin{table}
\caption{
Results for the networks corresponding to
the four programs we have studied. $N$ is the total number of nodes;
$\mu$ is the average number of links per
node; $C$ is the clustering coefficient, and $C_{rand}$ is
its value for an equivalent random network; $L$ is the average
shortest path, and $L_{rand}$ is the same quantity for the
corresponding random network.}
\begin{center}
\begin{tabular}{|c|c|c|c|c|c|c|}    \hline
\emph{program}  &
$N$  &  $\mu$  &
$C$  &  $C_{rand}$  &
$L$  &  $L_{rand}$ \\ \hline
Linux kernel   &
1448  &  41.4  &
0.88  &  0.03  &
2.11  &  1.93  \\ \hline
Mozilla        &
3803  &  76.6  &
0.81  &  0.02  &
2.49  &  1.87  \\ \hline
XFree86        &
1465  &  33.0  &
0.81  &  0.02  &
2.56  &  2.05  \\ \hline
Gimp           &
403   &  43.9  &
0.83  &  0.11  &
2.28  &  1.56  \\ \hline
\end{tabular}
\end{center}
\end{table}

A random network with given $N$ and $\mu$ (with $N\gg\mu$) is
characterized by having small values of $L$ and $C$. In particular,
for $N\rightarrow\infty$ and $\mu$ fixed, the average shortest path
in the largest connected component
approaches the logarithmic behavior of a Moore graph \cite{Bollobas:book},
\begin{equation} \label{eq:L_rand}
L_{rand} \approx \frac{\ln{N}}{\ln{\mu}}, 
\end{equation}{
and the clustering coefficient approaches zero \cite{WS:1998},
\begin{equation} \label{eq:C_rand}
C_{rand} \approx \mu/N.
\end{equation} 
On the other hand, regular networks are typically highly
clustered, but at the price of having very large $L$.
Small-world networks lie in between these two extremes.
They have large clustering, $C\gg C_{rand}$,
and small average shortest path, $L\approx L_{rand}$,
where $C_{rand}$ and $L_{rand}$
are the respective statistical quantities for a random network with the
same parameters $N$ and $\mu$.
From Table I, we see that the networks
corresponding to all four programs we have studied are small-world
networks. This result seems to be typically true for any large enough program.
Therefore, we conclude that the logical structure of large
programs can be described by small-world networks.

Notice that each source file corresponds to a totally connected
subgraph in the network, since every header file included in a source
file is connected to every other header file included in that same
source file. Thus the network consists of several clusters
(corresponding to the source files) interconnected by header files
that are included in more than one source file. The clustering effect
of the source files is the same as movies in the actors' network (the
``Kevin Bacon network''). Because of this, it is perhaps not
surprising that $C$ is large for our program networks. The fact that
$L$ is small, however, is not obvious and is due to nodes between otherwise distant clusters,
caused in turn by header files included in more than one source file.

\begin{table}
\caption{
Results for the networks constructed from the ones used in Table I by
deleting all the  nodes with a number of links larger than $N/4$.}
\begin{center}
\begin{tabular}{|c|c|c|c|c|c|c|}    \hline
\emph{program}  &  
$N$  &  $\mu$  &  
$C$  &  $C_{rand}$  &
$L$  &  $L_{rand}$ \\ \hline 
Linux kernel   &
1397  &  20.8  &
0.85  &  0.01  &
2.85  &  2.34  \\ \hline
Mozilla        &
3760  &  68.0  &
0.80  &  0.02  &
2.72  &  1.93  \\ \hline
XFree86        &
1435  &  30.8  &
0.80  &  0.02  &
2.79  &  2.09  \\ \hline
Gimp           &
241   &  24.9  &
0.74  &  0.10  &
2.55  &  1.66  \\ \hline
\end{tabular}
\end{center}
\end{table}

We have also investigated the influence of very highly connected nodes
on the network, and how the  networks' statistical properties
change if those highly connected nodes are removed. In order to do
this, we define a new network from each of the four original programs by removing all the 
nodes with a number of links larger than $N/4$. The new networks will,
of course, have smaller $N$ and $\mu$, and a larger $L$. We now
calculate $C$ and $L$ for these new networks. The results are
displayed in Table II. We see that these networks still have the
small-world property, in all cases. In fact, we have verified that the 
further removal of highly connected nodes always preserves the small-world
property of the resulting networks, up to the point where we
remove too many nodes, and the resulting networks are too small to
define meaningful statistics. This shows that the small-world
property in these networks is a robust phenomenon, and does not depend
on the presence of a few highly connected nodes in the tail of the algebraic
distribution (\ref{eq:algebraic}).

Finally, we observe that a network that contains full information
about both header and source files can be defined.  The result is a
bipartite network \cite{newman}, which has two types of nodes (one
corresponding to header files and the other to source files) and links
that run only between nodes of different kinds, as defined by the
``$\#$include'' clause.  The networks analyzed so far correspond to
the projection of this bipartite network onto the space of header
files.  A similar projection with respect to the space of source files
produces a network, whose nodes are source files and links are between
source files that include a common header file. The network of header
files and the network of source files share similar properties.  In
particular, both evolve according to a preferential growth and both
exhibit the small-world feature.

In summary, we have shown that large computer programs correspond to
growing networks that generally possess the small-world and
scale-free properties.  As computer softwares for various modern
applications are becoming increasingly more complex, it is important
to study and understand their topological structure for improved
efficiency and improved performance.  In particular, even for large
computer programs the flow of information within the program is
expected to be quite efficient because, as we have shown, in spite of
the size of the program the average shortest path in the underlying
network is very small.  Also, some of the nodes of these networks appear
to be much more connected than the average, which means that the
corresponding files in the program are required for a large number of
applications, making them relatively more important. This in turn,
together with the very fact that different parts of the program
(different applications) make use of a limited number of files, is
expected to help the maintenance and debugging of the programs. In
debugging, for example, the first files to be checked should be the
most connected ones.  We emphasize that our viewpoint that
sophisticated computer softwares can be considered as networks is
relevant because the network features identified in this paper are
expected to be generic and universal.

This work was supported by the AFOSR CIP (Critical Information
Protection) Program under Grant No. F49620-01-1-0317.

\end{document}